\documentclass[12pt]{article}
\usepackage[english]{babel}

\usepackage{graphicx}
\usepackage{color}
\usepackage{cite}

\begin{document}
\author{S.V. Poltavtsev, I.I. Ryzhov, V.S. Zapasskii, and G.G. Kozlov}
	%\affiliation{Spin Optics Laboratory, St.-Petersburg State University, 1 Ul'anovskaya, Peterhof, St.-Petersburg 198504, Russia}

\title{Noise spectroscopy of the optical microcavity:\\
nonlinear amplification of the spin noise signal\\ and giant noise}
\maketitle

\begin{abstract}
The spin-fluctuations-related Kerr rotation noise of the optical beam reflected from a microcavity with a quantum well in the intermirror gap is studied. In the regime of anti-crossing of the cavity polariton branches, the several hundred times enhancement of the noise signal, or giant noise, is observed. The effect of the amplification of the noise signal is explained by the nonlinear instability of the microcavity. In the frame of the developed model of built-in amplifier, the non-trivial properties of the noise signal in the regime of the negative detuning of microcavity are described.
\end{abstract}
 
\section*{ Introduction }

The study of spin fluctuations by means of optical polarimetry -- optical spin noise spectroscopy -- becomes more and more popular during the last years \cite{Zap1}. As it was first demonstrated in \cite{Zap}, the spin fluctuations in atomic system cause the fluctuations of its gyrotropy, which can be detected by means of a sensitive optical polarimeter as a noise of rotation of the polarization plane of the probe beam (Faraday rotation noise). The frequency spectrum of the observed noise is connected with the frequency dependence of the magnetic susceptibility by the fluctuation-dissipation theorem and in the simplest case has a Lorentzian shape with its width defined by the spin relaxation time and centered at the Larmor frequency \footnote{ proportional to the transverse static magnetic field applied to the studied system in Voight geometry}. Since the sample under study in the experiments is, as a rule, transparent for the probe beam, the optical noise spectroscopy can be considered as non-perturbative method for observation of the magnetic susceptibility. 

The growth of popularity of the optical noise spectroscopy during the last years is associated with the usage of the fast digital spectrum analyzers, which allow one to detect the magneto-spin noise of atomic and semiconductor systems in the radio-frequency (RF) range \cite{Crooker,Os,Mitsui}. 
 
For the optical noise spectroscopy experimental set-up based on RF spectrum analyzer, several improvements increasing the sensitivity were suggested \cite{Zap2,Mitsui1}. Besides, the new method for the spin noise detection was suggested in \cite{Os1}, which allows one to observe the spin noise in microwave frequency range. 

The noise spectroscopy of semiconductor systems features its own specificity. The up-to-date technologies of epitaxial growth make it possible to produce the semiconductor systems with precisely controllable spectral properties - quantum wells (QW) and quantum dots. 

Recently, the observations of the spin noise in these systems were reported \cite{Os2,Glazov}, so, at present one can speak about the noise spectroscopy of the nano-objects. Placing such an object into the intermirror gap of a planar Bragg microcavity allows one to increase drastically the noise signal observed as a noise of Kerr rotation of the optical beam reflected from such a complex system \cite{Glazov}. This gives the grounds to believe that optical noise spectroscopy of semiconductor nano-objects in microcavities becomes a part of the physics of the microcavities \cite{Kavokin}.
 
In the present paper, we study the properties of the Kerr rotation noise of the optical beam reflected from (Al,Ga)As Bragg $\lambda$-microcavity with GaAs QW at the center of the intermirror gap. 
 We use the sample with  spatial gradient of specral position of the cavity photon mode described in \cite{Rapaport2001,Zap3}. 
  In the region of the sample where the frequency of the cavity photon mode is well below that of QW resonances (region  of negative detuning) the RF noise spectrum of the Kerr rotation displays non-trivial bimodal shape described in \cite{Glazov}. The most unusual behaviour relates to  the  region, where the cavity photon mode frequency is close to the QW exciton resonance (region of anti-crossing), where the several hundred times increase of the noise signal, the {\it giant noise}, is observed. In our paper we show that the above properties of the noise signal can be explained by the nonlinear instability of the microcavity giving rise to the increase of the sensitivity of the reflectivity coefficient to the noise fluctuations of the refraction index within the intermirror gap.

The paper is organized as follows. In the first section, the experimental set-up, the sample under study and the results obtained are described. In the second section, the hypothesis of a built-in amplifier is formulated for the explanation of the giant noise observed in our experiments. On the basis of this hypothesis the formula describing the bimodal noise spectrum in the regime  of negative detuning is obtained. In the third section, all observed properties of the sample under study are qualitatively explained by the suggested simple model of the instable nonlinear resonator. All the results obtained in the paper are briefly summarized in Conclusion.

\section{Experiment}

We study the sample representing the planar Bragg $\lambda$-microcavity with GaAs QW (102/200/102 \AA AlAs/GaAs/AlAs) placed in the center of the intermirror gap (see \cite {Glazov,Rapaport2001} for the sample details). Note that in our experiments the strong coupling regime described in \cite{Rapaport2001} was partly violated by optical nonlinearity (see \cite{Zap3} for details). Experimental set-up used in our study is build for the observation of the Kerr rotation noise and described in detail in \cite{Glazov}, therefore we will restrict ourselves by a brief reminding. The linearly polarized monochromatic probe beam from tunable Ti:sapphire laser was focused to $\sim 20$ $\mu$m spot of the sample in normal direction. The sample was placed in the closed cycle cryostat and cooled down to $3\div 5$ K. Electro-magnet was used to apply external transverse magnetic field on the sample. The beam reflected from the sample was directed to a polarimetric detector comprised of phase plate, polarizing beam splitter (Wollaston prism), and broadband ($\delta\nu$ = 200 MHz) balanced photodetector.

The fluctuations of the magnetization in the intermirror gap (related to the spin fluctuations of free carriers or excitons in QW) cause fluctuations of polarization of the reflected beam and can be detected as a noise electric signal on the output of the photodetector. The wavelength of the probe beam was tuned to maximize the amplitude of the noise signal.

A typical experiment on the spin-noise spectroscopy \cite{Zap,Crooker,Os,Mitsui} implies measurement of the RF-spectrum of this signal as a function of magnetic field strength. This spectrum is detected using a spectrum analyzer and, in the simplest case, has a Lorentzian shape with the peak position shifting linearly with magnetic field strength. From the reflectivity spectra of the studied sample (presented in \cite{Rapaport2001}) it is seen that the spectral position of the cavity photon mode reveals  a strong spatial dependence. Bearing this in mind, it is convenient to specify the following three regions of the sample:

\noindent(1) {\it The region of negative detuning}, where the frequency of the cavity photon mode is below those of QW resonances. The region of negative detuning corresponds to the spatial interval on the sample surface $ x\in $ [0, 0.3] mm with respect to the sample edge (see \cite{Zap3} for details). 

\noindent(2) {\it The region of anti-crossing of the polariton branches}, where the frequency of the cavity photon mode "passes" through the frequencies of QW resonances ($x\in$ [0.3, 0.9] mm).

\noindent(3) {\it The region of positive detuning}, where the frequency of the cavity photon mode becomes higher than those of QW resonances ($x> 0.9$ mm). In this region the noise signal is strongly suppressed, because in this case  the used asymmetric microcavity operates in  so called post critical regime,  described in \cite{Zap3}.

The properties of the noise signal in the first two regions are essentially different and sequentially described below.

{\it The region of negative detuning.} The Kerr-rotation-noise power spectra measured in this region of the sample at different Voight magnetic fields are presented in Fig.1 (noisy curves). These spectra exhibit the non-trivial bimodal shape at non-zero magnetic fields: besides the ordinary spin noise peak at Larmor frequency, whose shift in magnetic field is defined by $g$-factor $|g|\approx0.35$, there is a well pronounced maximum at zero frequency. The amplitude of this maximum decreases with rising magnetic field, while its spectral position does not depend on the field strength. Dependence of the noise spectra on the probe beam intensity displays noticeable deviation from quadratic growth, which corresponds to the optical nonlinearity. The amplitude of the zero-frequency maximum in the noise spectrum demonstrates the most pronounced nonlinear behavior: the ratio between the amplitude of this maximum and that of the maximum at Larmor frequency (at fixed magnetic field) diminishes with decrease of the probe intensity. The typical value of the noise signal in this sample region is of the order of the shot noise  of the probe whose intesity in our experiments $\sim 1$~mW. 	 

\begin{figure}
	\begin{center}
		\includegraphics[width=.8\columnwidth,clip]{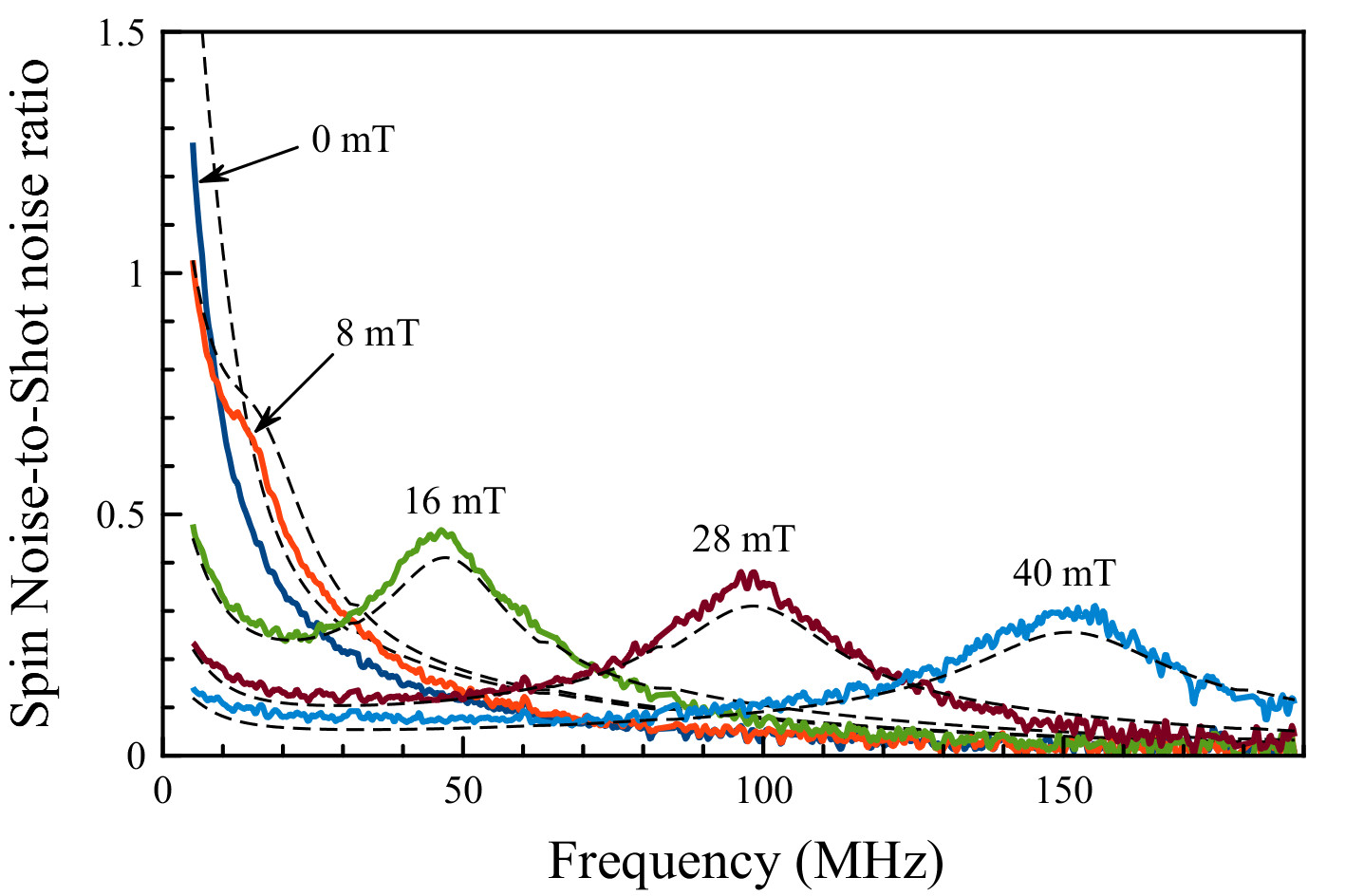}
		\caption{The noise spectra in the region of the negative detuning at different magnetic fields. {\it Noisy curves} -- experiment, {\it smooth dashed  curves} -- fitting by Eq. (\ref{11}). }
		\label{1}
	\end{center}
\end{figure}

\begin{figure}
	\begin{center}
 				\includegraphics[width=.8\columnwidth,clip]{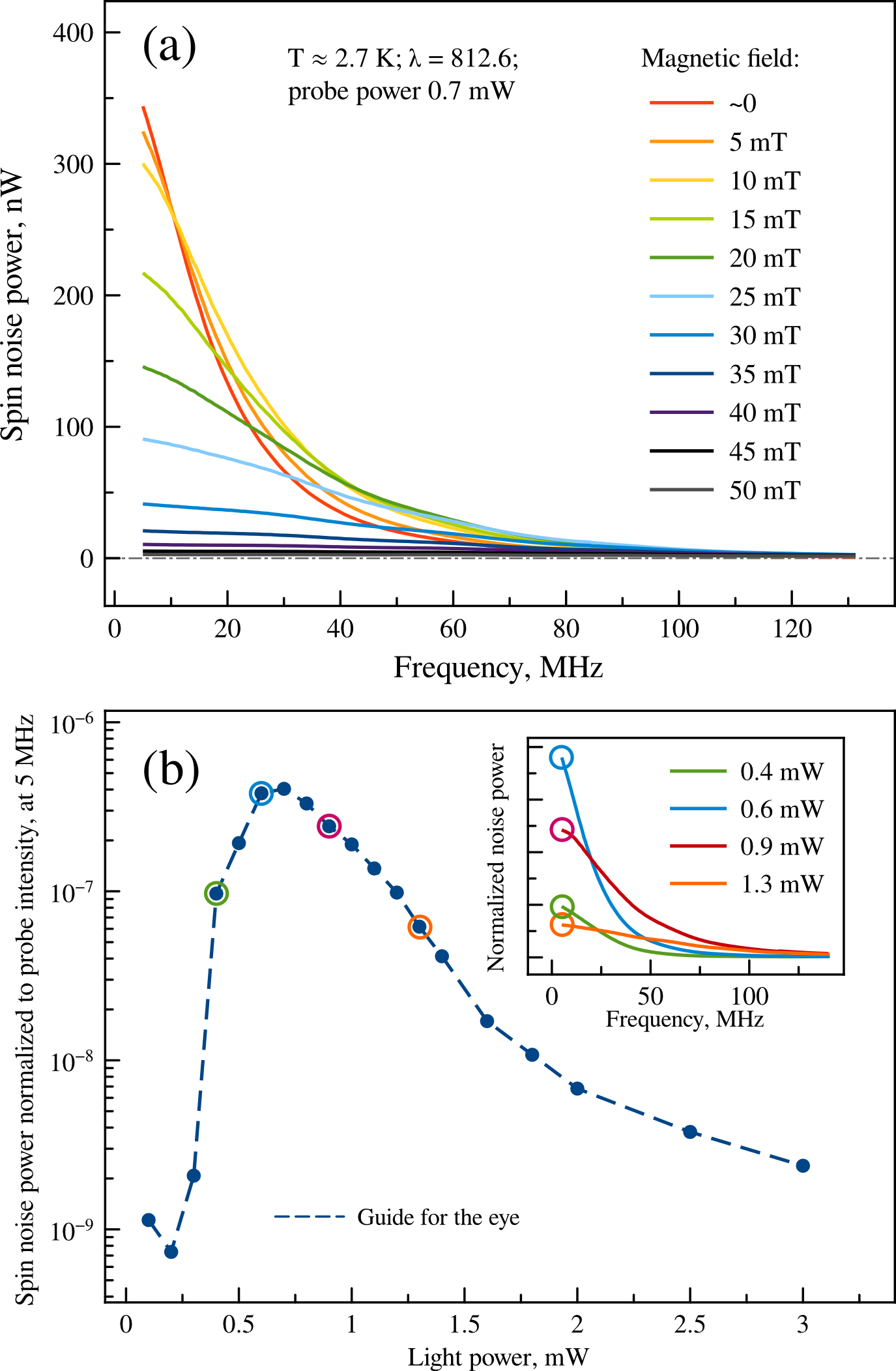}
 			\caption{(a) spectrum of giant noise at different magnetic fields. (b) the amplitude of the spectrum of giant noise at zero magnetic field at various intensities of the probe beam. Inset: noise spectra in the region of giant noise for various intensities of the probe.}
 		\label{2}
	\end{center}
\end{figure}

{\it The region of polariton branches anti-crossing.} By varying the probe spot around the anti-crossing region it was possible to locate the areas of {\it giant noise}, where the amplitude of the noise signal at zero magnetic field was  several hundred times larger than that in the region of negative detuning. The spectrum of the giant noise has the form of the peak at zero frequency with width $\sim 10\div20$ MHz, whose amplitude  reduced with increasing of  magnetic field (Fig.\ref{2}a). The amplitude of the giant noise peak reveals the essentially nonlinear behavior with  varying of probe intensity: steep increase at moderate intensities of the probe and reduction at  the intensities $>1$ mW  (Fig.\ref{2}b). 

The above spectra were measured by means of spectrum analyzer in the RF-range with lower limit of 5 MHz. The direct observation of the output of the polarimetric detector by means of the oscilloscope revealed that the giant noise is accompanied by the chaotic low frequency oscillations of the reflected beam polarization with characteristic frequencies $\sim 1 - 50$ kHz. The mentioned instability appeared  at strong enough intensities of the probe. The intensity of the reflected beam exhibited similar behavior.

 \section{The hypothesis of the built-in amplifier}
 
The results presented in the previous section unambiguously show the important role of the optical nonlinearity in the formation of the Kerr rotation noise signals observed in our experiments. The nonlinear behavior of the optical cavities is well known and studied in detail \cite{Gibbs,Gibbs1,McCall}. 
  In our case, the nonlinearity is caused not only by the resonant increasing of the electromagnetic field in the microcavity, but also by using of the sharp focusing of the probe beam. 

\begin{figure}
	\includegraphics[width=.8\columnwidth,clip]{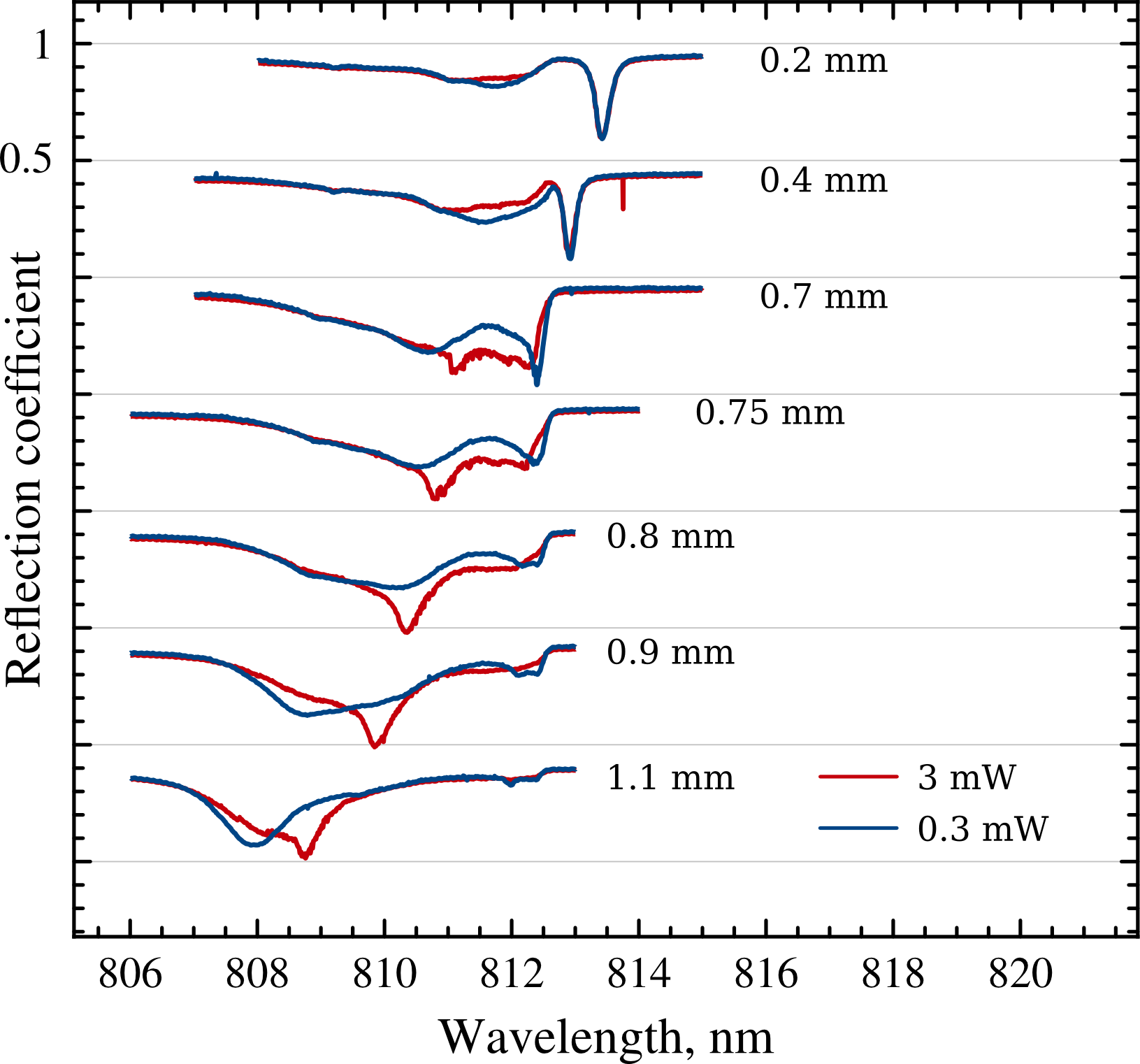}
	\caption{ The reflectivity spectra of the studied microcavity with QW at different relative spectral position of the cavity photon mode and  QW resonances. The spectra 0 -- 0.3 mm correspond to the region of the negative detuning; spectra 0.3 -- 0.9 mm – to the region of anti-crossing of the polariton branches.  Measurements were carried out for the probe beam intensities 3 mW ({\it red curves}) and 0.3 mW ({\it blue curves}).}
	\label{3}
\end{figure}

The reflectivity spectra measured at different positions  of the probe
 (in mm with respect to the sample edge)
 at the sample surface for two probe intensities, 0.3 mW and 3 mW, are shown in Fig.\ref{3} and allow one to judge about the degree of optical nonlinearity. The observation of reflectivity spectra Fig.\ref{3} and the detection of the noise spectra were performed under similar conditions.

It is seen from the Fig.\ref{3} that the influence of the increase of the probe beam intensity on the reflectivity spectra is well-pronounced even in the region of the negative detuning and becomes dramatic in the region of anti-crossing of the polariton branches, where the giant noise is observed. This behavior of the reflectivity spectra allows us to suggest the following qualitative explanation of the oscillatory instability of the microcavity described in the previous section. 

After switching on the resonant probe beam, the amplitude of oscillations of the optical field in the microcavity starts to grow. Due to the nonlinearity of the medium in the intermirror gap~\footnote{related, for instance, to the saturation of the QW exciton resonance}, the microcavity can tune out of the resonance. This leads to the decreasing of the amplitude of the optical field in the microcavity and, consequently, it tunes back to the resonance and the amplitude of the optical field starts to grow again, so, this process occurs repeatedly. The similar scenario of self-excitation is described in \cite{Gibbs}.

This (or similar) mechanism of self-excitation (i.e. appearance of the oscillations of the amplitude of the field) of the microcavity must exhibit a threshold behavior as a function of the probe beam intensity. It is well known that the similar systems are very sensitive to the variation of their parameters at the threshold of self-excitation. For this reason, it is naturally to expect the dramatic growth of the sensitivity of the reflection coefficient~\footnote{ and the polarization of the reflected radiation either} of the nonlinear microcavity to the fluctuations of the optical properties of the medium in the intermirror gap when the probe intensity is close to the self-excitation threshold. This is confirmed by the analysis of the simple model presented in the next section.

{\it In our opinion, the appearance of the giant noise from the regions of the sample with strong optical nonlinearity is related to the pre-threshold growth of the sensitivity of reflection coefficient of the nonlinear microcavity to spin fluctuations~}\footnote{exactly, to the fluctuations of the gyrotropy}.

Based on this hypothesis, we can write the following expression for the noise signal $S$ corresponding to the fluctuations of the Kerr rotation: $S=(\xi+1)M$, where $M$ is the noise signal from a hypothetical {\it linear} microcavity and $\xi$ -- is a factor accounting the amplification of the polarimetric response of the microcavity to the fluctuations of optical properties of the medium in the intermirror gap related to the  medium nonlinearity. In agreement with the aforesaid, the factor $\xi$ should depend on the intensity $I$ of the probe $\xi=\xi(I)$, with $\xi(I=0)=0$ and, when $I$ is close to the threshold of self-excitation $I_c$, it must be $\xi(I\rightarrow I_c)\gg 1$. Therefore, one can say that nonlinearity of the microcavity leads to the appearance  of some {\it built-in amplifier} of its polarimetric response with amplification factor $\xi(I)$. Since in our experiments the dependence of the noise power on frequency $\omega$ is observed, one should take into account the possible frequency dependence of the amplification factor: $\hskip3mm\xi=\xi(I,\omega)$.

As it was mentioned in the previous section, the noise signal from our sample exhibit noticeable nonlinear character even in the region of the negative detuning where no giant noise is observed. In what follows, we show that the hypothesis of the built-in amplifier naturally explains the presence of the  zero-frequency maximum   in the noise spectrum described in the first section. To do this, we assume that the frequency dependence of the amplification factor $\xi$ has the shape of the zero-centered Lorentz curve with a width $\Delta$: \footnote{the simple model described in the next section justifies the dependence of this type } $\xi(I,\omega)=\xi_0(I)\Delta\pi {\cal L}(\Delta,\omega)$, where ${\cal L}(\Delta,\omega)\equiv \pi^{-1}\Delta/[\Delta^2+\omega^2]$, and $\xi_0=\xi(I,0)$. Then, in accordance with the hypothesis of built-in amplifier, one can write down the following expression for the noise spectrum observed in our experiments:

\begin{equation}
	S(\omega)=\bigg (\xi_0(I) \pi\Delta \hskip2mm {\cal L}(\Delta,\omega) +1\bigg )
	\bigg [ {\cal L}(\Delta_e+\ae h,\omega-g \beta h/\hbar)+
	{\cal L}(\Delta_e+\ae h,\omega+g \beta h/\hbar) \bigg ]
	\label{11} 
\end{equation} 
 
\noindent Here, the factor in the square brackets corresponds to the classical spin noise spectrum in the magnetic field $h$: the Lorentzian peak with half-width at half-maximum $\Delta$ centered at Larmor frequency $g\beta h/\hbar$. \footnote{In some regions of our sample the spin noise spectrum has the shape close to the ''triangular'' } The parameter $\ae$ describes the noticeable broadening of the noise spectrum in the large magnetic field observed in our experiments (this effect can be explained by small dispersion of $g$-factors of the spins responsible for the noise signal). The results of  fitting of the experimental noise spectra by Eq. (\ref{11}) are presented in Fig.\ref{1} (smooth curves). The values of the fitting parameters are: $\xi_0=2.5, \Delta= 2\pi \times 7$ MHz, $\Delta_e=2\pi\times 20$ MHz, $\ae=2\pi\times 0.3$ MHz/mT$, g=0.35$.

As it was expected, the amplification factor $\ae$ appeared to be small, as the optical nonlinearity in the considered region of negative detuning is small. In our case, $\xi_0\approx 2$, which leads to the bimodal shape of the observed noise spectrum: the amplitude of the zero-frequency peak in the noise spectrum (corresponding to the low-frequency spectral components of spin noise \footnote{whose spectrum has maximum at the Larmor frequency} amplified by the built-in amplifier) and the amplitude of the maximum at Larmor frequency have the same order of magnitude. Upon decreasing the probe intensity, the amplification factor $\xi$ (governed by the optical nonlinearity) is decreasing. This leads to the observed suppression of the zero-frequency maximum in the noise spectrum. In the regions of the sample possessing larger optical nonlinearity (the regions of anti-crossing of the polariton branches), where the giant noise is observed, the amplification factor grows and in the spots of giant noise has an order of several hundreds. The Eq. (\ref{11}) shows that in this case the observed signal represents only the spin noise amplified by the built-in amplifier –- the unity in round brackets in Eq.(\ref{11}) can be neglected. Due to the fact that the effective amplification takes place only for the spectral components with frequencies $<\Delta$, only the spin noise components with frequencies $<\Delta$ are observed and spectrum of the giant noise has monomodal shape (Fig.\ref{2}). When magnetic field becomes non-zero, the amplitude of these components becomes lower because the maximum of the spin noise spectrum (at the Larmor frequency) is shifting towards high frequencies. This explains the observed reduction of the giant noise in magnetic field.

%\newpage 
\section{The model of nonlinear oscillator }

The development of the detailed model of nonlinear oscillator (which take into account the concrete mechanism of optical nonlinearity, spatial inhomogeneity of the microcavity and focusing of the probe beam) at the present stage of the research seems to be premature. In this section, we develop the simplest model of instable nonlinear resonator exhibiting strong enhancement of sensitivity of its reflectivity to the fluctuations of the refractive index of the intracavity medium at the threshold of the self-excitation. Despite the fact that in our experiments we observe the {\it polarization properties of the reflected beam,} the below simple model  reveals, in our opinion, the most important {\it qualitative} properties of our microcavity and to the considerable extent justifies the suppositions of the above hypothesis of built-in amplifier. 

Let's consider an effective oscillator with the eigen frequency $\omega$ and the width of resonance  $\Delta\omega$ and put it into the correspondence to our microcavity. Then, the amplitude $A$ of the field oscillations in the microcavity is given by the relationship:

\begin{equation}
	A= {\imath E \hskip1mm \Delta\omega \sqrt{Q}\over \omega_0-\omega+\imath\Delta\omega}
	\hskip10mm Q\equiv {\omega\over \Delta\omega},
	\label{21}
\end{equation} 

\noindent where $E$ and $\omega$ -- are amplitude and frequency of the optical oscillations in the incident beam, and quantity $Q$ (at $\omega\sim\omega_0$) is the resonator Q-factor (we assume $Q\gg 1$). The amplitude $R$ of the wave reflected from the cavity, can be presented as the sum of the amplitude $r_tE$ of the wave corresponding to the probe reflected by the top mirror ($r_t$ -- is the reflectivity coefficient of the top mirror) and the amplitude $tA$ of the wave escaped from the cavity through the top mirror ($t$ -- is the transmission coefficient of the top mirror, $|t|\ll 1$): $R=r_tE+tA$. Using Eq. (\ref{21}) one can express $t$ via the cavity resonant reflection coefficient $r_{res}=R/E|_{\omega=\omega_0}$, which can be easily estimated in the experiment and for the high Q-factor symmetrical cavities is a small real number $|r_{res}|\ll 1$. After that, the following relationship for the reflected field can be obtained: $R=r_tE+(r_{res}-r_t) A/\sqrt Q$. 
 
The fact that the refractive index of the intracavity medium is dependent on the intensity $|A|^2$ of the intracavity optical field, can be the reason for the optical nonlinearity of the microcavity under study. In the simplest case, this fact can be taken into account by assuming that the eigen frequency of the cavity $\omega_0$ depends on the intensity of the optical field $|A|^2$. In our model, we consider this dependence to be non-local (retarding) and consisting of two contributions: the fast one $\Omega_f$ (with characteristic time $T_f$) and the slow one $\Omega_s$ (with characteristic time $T_s\gg T_f $): \footnote{The similar mechanism of optical nonlinearity leading to the oscillating regime of GaAs-based optical resonator, was described in \cite{Gibbs1,McCall}}

\begin{equation}
 \cases{
 \omega_0=\bar\omega_0+\Omega_s+\Omega_f \cr
 \dot\Omega_{s(f)}+\Omega_{s(f)}/T_{s(f)}=\nu_{s(f)} |A|^2}
 \label{31}
\end{equation}

\noindent Here, $\bar\omega_0$ is the eigen frequency of the cavity in linear regime, $\nu_s$ and $\nu_f$ -- are constants describing above contributions and related to the cavity nonlinearity. If we measure   time  in units of $\Delta\omega^{-1}$,  frequencies -- in units of $\Delta\omega$, and if the amplitude of the intracavity field is defined by the dimensionless quantity $a\equiv A/\sqrt Q E$, then, using Eqs. (\ref{21}) and (\ref{31}) one can obtain the following set of equations describing the dynamics of our nonlinear resonator:
 
\begin{equation}
 	\cases{ [1+\imath z]a=1\cr
	z=b-\theta_s-\theta_f\cr
	\dot \theta_{s(f)}+\theta_{s(f)}/\tau_{s(f)}=g_{s(f)}|a|^2}
	\hskip5mm g_{s(f)}={\nu_{s(f)} QE^2\over \Delta\omega^2}
	\hskip5mm \theta_{s(f)}\equiv \Omega_{s(f)}/\Delta\omega\hskip5mm
	b\equiv {\omega-\bar\omega_0\over \Delta\omega},\hskip5mm
	\label{41}
\end{equation} 
 
\noindent where $\tau_{s(f)}\equiv \Delta\omega T_{s(f)}$, and the quantity $z=[\omega-\omega_0]/\Delta\omega$ corresponds to the $|a|^2$-depending dimensionless detuning of the cavity. If the parameters of the cavity acquire some variations, the amplitude of the field in the cavity reaches its steadystate value with the characteristic time $\sim \Delta \omega^{-1}$, which we assume to be unit. Therefore, the set of Eqs. (\ref{41}) makes sense under the condition $\Delta\omega^{-1}\ll T_{s(f)}$, which we will assume to be satisfied. 
 
Let's now show that the above model of nonlinear cavity describes the spontaneous oscillations of the field amplitude $a$ in the cavity. For this reason, we consider the behavior of our cavity at the time scales much greater than $T_f$. Under this assumption, one can consider the fast contribution $\Omega_f$ to be instantaneous and write $\theta_f=c|a|^2$ (where $c\equiv g_f \tau_f$). After that the set of Eqs. (\ref{41}) can be   reduced to the form

\begin{equation}
 \cases{ [1+ z^2]|a|^2=1\cr
 z=b-\theta_s-c|a|^2\cr
 \dot \theta_{s}+\theta_{s}/\tau_{s}=g_{s}|a|^2}
 \label{51}
\end{equation}

\begin{figure}
\begin{center}
\includegraphics[width=.8\columnwidth,clip]{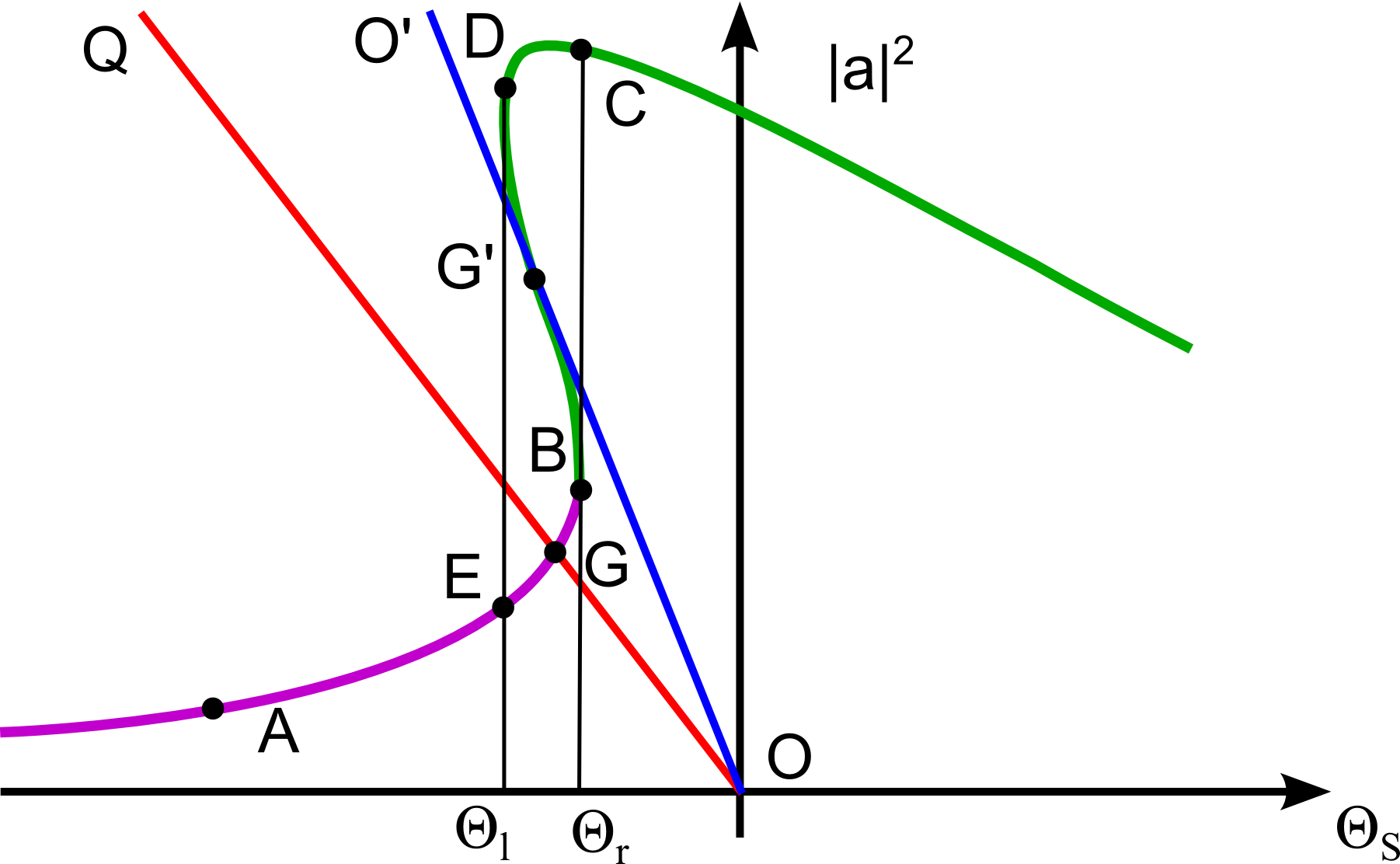}
\caption{S-shaped dependence }
\label{4}
\end{center}
\end{figure}

From the first two equations, one can express $|a|^2$ via $\theta_s$. Since the dependence $|a|^2(\theta_s)$ is obtained by solution of a cubic equation, it can be multivalued function. This can be shown by graphical analysis of equation for the detuning $z$ (which can be obtained making use of the relationship  $|a|^2=[1+z^2]^{-1}$), whose solutions correspond to the crossing-points of the Lorentz curve $y=c/[1+z^2]$ and a line $y=b-\theta_s-z$. If the amplitude $|c|$ of the lorentzian is small enough,  the equation for the detuning has the single real root $z_1\approx b-\theta_s$. Otherwise, it can be shown that if $|c|>c_{cr}=8\sqrt 3/9$ there appears an interval of  $\theta_s$, for which the equation for the detuning has three real roots $z_1,z_2,z_3$. Therefore, at $|c|>c_{cr}$ (below we will consider $c>0$) the dependence $|a|^2(\theta_s)$ becomes multivalued and in the above interval of $\theta_s$ has the form of S-shaped curve (see Fig. \ref{4} where the interval of ambiguity is $\theta_s\in [\theta_l,\theta_r]$). At decreasing $|c|$, the region of ambiguity decreases and vanishes at $|c|=c_{cr}$: $\theta_l=\theta_r |_{c=c_{cr}}$. Using the dependence $|a|^2(\theta_s)$ presented at Fig.\ref{4}, one can completely describe the dynamics of considered model of nonlinear cavity. Each point of the plain $(|a|^2, \theta_s)$ maps the state of the cavity at an arbitrary time moment.  For actuall dynamics of the nonlinear microcavity this point must  belong to the S-shaped curve (Fig.\ref{4}) -- in this case the first two equations of the set (\ref{51}) are satisfied. The character of motion of the mapping point {\it along} this curve is defined by the third equation of the set (\ref{51}) and represents the moving of the mentioned point towards the crossing-point of the $S$-shaped curve and a  line $|a|^2=\theta_s/g_s\tau_s$. This point corresponds to the stationary state of the system.

Two essentially different types of motion are possible. The first type takes place if the position of line $|a|^2=\theta_s/g_s\tau_s$ is similar  to that of line $OQ$. In this case the mapping point moving from the starting point $A$ along the $S$-shaped curve to the right will reach  the point $G$, where it will stay infinitely long. The second type of motion takes place if the parameters of the system are such that the line $|a|^2=\theta_s/g_s\tau_s$ crosses the $S$-shaped curve at some point $G'$ within the {\it returning } fragment $BD$ of the multivalued interval $[\theta_l, \theta_r]$  (for example as the line $OO'$ at Fig.\ref{4}). In this case, the mapping point moving from the point $A$ along the $S$-shaped curve towards the point of stationary state $G'$ will come to the point $B$, from which the system will jump to the point $C$. After that, the mapping point will proceed its movement towards the point $G'$ (corresponding to the stationary state of the system) along the fragment $CD$. At the point $D$ the system, again, will jump to the point $E$, and so on. Therefore, in the case of the second type of motion, the mapping point moves periodically along the loop $EDCB$ consisting of two verticals $ED$ and $CB$ and of two fragments $EB$ and $DC$ of the $S$-shaped curve. It is seen from Fig.\ref{4} that in order to gain the oscillatory regime at $c>0$, the condition $g_s<0$ must hold.

The third equation of the set (\ref{51}) can be written in the form $\dot\theta_s =g_s \hskip1mm |a|^2(\theta_s)-\theta_s/\tau_s $. Taking this into account, one can write the following relationships for the dimensionless time intervals $\tau_{EB}$ ($\tau_{DC}$), during which the mapping point passes the fragment $EB$ ($DC$), and for the total period $\tau_A$ of oscillations: 
 
 \begin{equation} 
 \tau_{EB (DC)}= \bigg |\int_{EB (DC)}{d\theta_s\over g_s |a|^2(\theta_s)-\theta_s/\tau_s}\bigg |
 \hskip10mm \tau_A=\tau_{EB}+\tau_{DC}
 \label{6}
 \end{equation}

The multivalued function $|a|^2(\theta_s)$ (Fig.\ref{4}) entering these relationships can be obtain either numerically or by means of Cardano formula. The direct calculations by Eq. (\ref{6}) have shown that the period of oscillations significantly depends on the intensity-related parameters of the problem, $c$ and $g_s$, and can be by order of magnitude smaller than $\tau_s$. Therefore, the above treatment is justified if $\tau_s$ exceeds $\tau_f$ more than by an order of magnitude.

Concluding the consideration of the oscillatory regime of the above model of nonlinear cavity, we itemize the main conditions of the self-excitation: (i) $|c|>c_{cr}=8\sqrt 3/9$. It means that the intensity $|A|^2$ of the field in the cavity must be high enough, so that nonlinear shift of the eigen frequency must be of the order of the resonance linewidth $\sim\Delta\omega$. \footnote{ Note that, as it is seen from for the reflectivity spectra in the anti-crossing region (Fig.\ref{3}), the similar condition (at least qualitatively) actually holds.} (ii) The line $|a|^2=\theta_s/g_s\tau_s$ (at $g_sc<0$) should cross the $S$-shaped curve within the multivalued region. Note that mutual position of mentioned line and $S$-shaped curve depends on the dimensionless detuning $b$. By varying parameter $b$ one can violate the self-excitation regime of the cavity.

Let's consider now the response of the above cavity to a small modulation of detuning $\delta b$ under condition of pre-threshold regime, when $c<c_{cr}$. In this case, the set of Eqs. (\ref{41}) has a stable stationary solution (we denote it by $a_0, z_0,\theta_{s0},\theta_{f0}$), which can be obtained from Eq. (\ref{41}) by omitting the terms $ \dot\theta_{s}$ and $\dot\theta_{f}$. The problem we are interested in can be formulated now in a following way: let's replace $b\rightarrow b+\delta b(t), \delta b/b\ll 1$ in Eq. (\ref{41}) and calculate the response $\delta r$ of the reflection coefficient of the cavity in the vicinity of stationary regime to the small  modulation of detuning $\delta b(t)$. In accordance with a standard procedure, one should present the sought solution of (\ref{41}) in the form $a(t)=a_0+\delta a(t), z(t)=z_0+\delta z(t), \theta_s(t)=\theta_{s0}+\delta\theta_s(t), \theta_f(t)=\theta_{f0}+\delta\theta_f(t)$, and find the quantities $\delta a, \delta z, \delta \theta_s, \delta\theta_f$ in the first order of the perturbation theory in $\delta b$. It is convenient to perform the calculation of the quantities $q\equiv$ 2Re~$[\delta a a^\ast_0]$ and $v\equiv$ 2Im~$[\delta a a^\ast_0]$, which are connected with the variation of the cavity reflectivity by means of relation $\delta r=(r_{res}-r_t)[q+\imath v]/2a_0^\ast$. If the temporal behavior of $\delta b$ represents harmonic oscillations with frequency $\nu$, then for the amplitudes of quantities $q$ and $v$ (we denote them as $q_0$ and $v_0$) one can obtain the following expressions: 

 \begin{equation}
|v_0|=|\chi (\nu)\delta b|\hskip5mm |q_0|=|z_0\chi (\nu)\delta b|,
\end{equation}

\noindent where susceptibility $\chi(\nu)$ is defined as

\begin{equation}
\chi(\nu)={2\over [1+z_0^2]^2-2z_0[g_f\tau_f/(1-\imath\nu\tau_f)+g_s\tau_s/(1-\imath\nu\tau_s)]}.
\label{8}
\end{equation}

The stationary detuning $z_0$ entering these formulas can be obtained from the equation

\begin{equation}
	b-z_0={g_f\tau_f+g_s\tau_s\over 1+z_0^2},
\end{equation} 

\begin{figure}
 	\includegraphics[width=.8\columnwidth,clip]{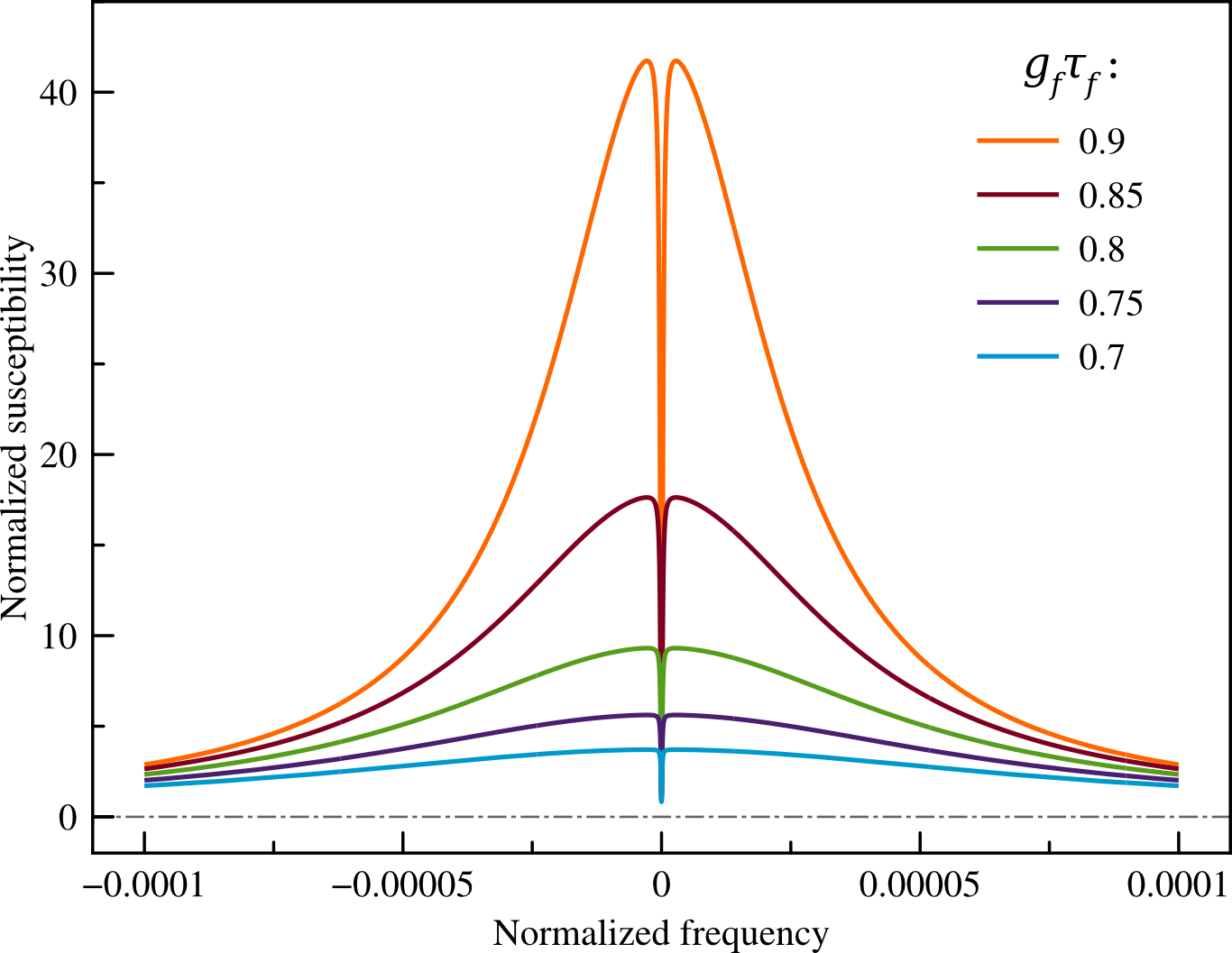}
 	\caption{Frequency dependence of the susceptibility (\ref{8}) 
	normalized to the linear susceptibility $|\chi(\nu,c)\hskip2mm (1+b^2)^2|^2$ 
	calculated for $c=g_f\tau_f = 0.7, 0.75, 0.8, 0.85,$ and $0.9$. 
 	Other parameters are: $b=1, \hskip1mm \tau_s=2\times 10^7,\hskip1mm \tau_f=0.0002\hskip1mm\tau_s,\hskip1mm  g_s=-0.6 \hskip1mm c/\tau_s$}
 	\label{5}
\end{figure}

\noindent which has the unique real solution in the stable regime of the cavity considered here. Using the Eq. (\ref{8}) for the susceptibility, one can calculate its behavior at the threshold of instability, $c\rightarrow c_{cr}$. The results of such calculation are presented in Fig.{\ref{5}}, which shows that the amplitude of the susceptibility $\chi(\nu)$ (and consequently the amplitude $\delta r$ of the reflection coefficient modulation) considerably grows, when $c\rightarrow c_{cr}$ with its shape being in qualitative agreement with that for the amplification factor (excluding the narrow dip at $\nu=0$). The widths of the main maximum and of the central narrow dip of the function $|\chi(\nu)|^2$ are defined by $\tau_f$ and $\tau_s$ 

Therefore, our model successively reproduces the main qualitative properties of the real microcavity, itemized at the beginning of this section. For this reason, the more detailed (and quite possible) analysis of the model is not required at present stage of the research and we restrict ourselves by two remarks only: (i) the central narrow dip of the function $|\chi(\nu)|^2$ is conditioned by the requirement $g_fg_s<0$, which is necessary for the cavity self-excitation. For this reason, this dip, probably, has a physical sense.\footnote{Measurement of the noise spectrum in the frequency region $< 2$ MHz in our experiments was hampered by the high level of the analyzer intrinsic noise in this region.} (ii) the assumption concerning the possibility of the slow and fast components of the response of refractive index to the probe intensity changes seems not to be fantastic: the fast part of this response can be associated with bleaching of excitonic susceptibility of the QW, and the slow part –- with the slow processes of redistribution of the photogenerated electrical charge.

\section*{Conclusion}

The properties of the Kerr rotation noise of the optical beam reflected from the Bragg microcavity with the QW in the intermirror gap are studied at various detuning between the cavity photon mode and the  QW resonances. In the region of anti-crossing of polariton branches, several hundred times enhancement of the noise signal, we call it {\it giant noise}, is discovered. This effect is explained on the basis of the hypothesis of built-in amplifier -– the pre-threshold enhancement of the polarimetric sensitivity of the nonlinear microcavity to the fluctuations of the gyrotropy in the intermirror gap. The properties of the noise signal in the region of negative detuning between the cavity photon mode and the  QW resonances are also explained in the frame of the built-in amplifier hypothesis.

Note that the described effect of nonlinear amplification, possibly, can be used for the enhancement of the sensitivity of the noise spectroscopy measurements. For this purpose, one should manage to control the parameters of the built-in amplifier, in particular, to make its frequency response homogeneous in the frequency range up to $\sim$ 200 MHz.

%\bibliographystyle{ieeetr}
%\bibliography{biblio}

\end{document}